\documentclass[a4paper]{article}

\usepackage[english]{babel}
\usepackage[utf8x]{inputenc}
\usepackage[T1]{fontenc}

\usepackage[a4paper,top=3cm,bottom=2cm,left=3cm,right=3cm,marginparwidth=1.75cm]{geometry}

\usepackage{setspace}
\usepackage{amssymb}

\usepackage{gensymb}

\usepackage{amsmath}
\usepackage{graphicx}
\usepackage[colorinlistoftodos]{todonotes}
\usepackage[colorlinks=true, allcolors=blue]{hyperref}

\title{Necessary conditions for out-of-plane lattice plasmons in nanoparticle arrays}

\author{Gordon Han Ying Li$^1$ and Guangyuan Li$^{2,*}$}

\date{}
\begin{document}
\maketitle

\begin{spacing}{2.0}

\noindent \large $^1$Institute of Photonics and Optical Science (IPOS), School of Physics, The University of Sydney, Sydney, New South Wales 2006, Australia

\noindent $^2$Shenzhen Institutes of Advanced Technology, Chinese Academy of Sciences, Shenzhen 518055, Guangdong Province, China

\noindent *gy.li@siat.ac.cn

\end{spacing}

\begin{abstract}
Out-of-plane lattice plasmons (OLPs) supported by metallic nanoparticle arrays are promising for diverse applications due to their remarkably narrow linewidths and significant enhancement of local fields. Here we investigate the necessary conditions for metallic nanoparticle arrays to achieve OLPs. Our results show that, besides the prerequisites of large height, oblique incidence, and TM polarization, which have been pointed out by the literature; the array period should be properly designed within a limited range, the dielectric environment should be homogeneous, and the height and the angle of incidence need to be optimized. We expect this work will advance the understanding and engineering of OLPs in metallic nanoparticle arrays and promote their applications in nanolasers, ultrasensitive biochemical sensors, and nonlinear optics.
\end{abstract}

\section{Introduction}
Plasmonic nanostructures have attracted increasing attention because they can concentrate light in small regions and thus greatly enhance light-matter interactions. One of the simplest plasmonic structures is a single metallic nanoparticle, which can support a localized surface plasmon resonance (LSPR) with large local field enhancement \cite{Pohl2005bowtieAntenna}. However, LSPRs suffer from short plasmon lifetimes (2-10 fs) \cite{Sayed1999SPRtime} and low quality factors ($Q<10$) \cite{Lilleyl2015LSPRQ}, which limits the local field intensity and the light-matter interaction enhancement. A powerful strategy to address this problem is to pattern nanoparticles in one- or two-dimensional arrays \cite{Schatz2016SLRreview,Odom2018SLRreview,Grigorenko2018SLRreview}. Due to strong coupling between neighboring nanoparticles, lattice plasmons supported by metallic nanoparticle arrays combine desirable photonic and plasmonic attributes, including suppressed radiative loss, high quality factor, and significant field enhancement over large volumes \cite{Schatz2016SLRreview,Odom2018SLRreview,Grigorenko2018SLRreview}. To date, most work has focused on in-plane lattice plasmons (ILPs) \cite{Schatz2004InPlaneSLRfirst,Kravets2008LatticeExp,Barnes2008LatticeExp,Crozier2008LatticeExp,Abajo2010ILPsub,Khlopin2017IPSLRAuAl}, which are produced due to inter-particle coupling of in-plane LSPR oscillations. It has been demonstrated that a homogeneous dielectric environment is beneficial for achieving narrow linewidth, although narrow linewidth can still be observed in an asymmetric dielectric environment provided that the grating is properly designed \cite{Grigorenko2014OLPLR,Khlopin2017IPSLRAuAl}. As a result, the effects of the array period, dielectric environment and nanoparticle dimensions on ILPs are well understood, and ILPs have already found attractive applications in nanolasers \cite{Odom2013IPSLRlasing,Odom2015IPSLRlasing}, nonlinear optics \cite{Matti2016PLR_SHG,Matti2018PLR_SHG,Halpin2018PLR_NLemission}, and ultrasensitive sensing \cite{Offermans2011SLRsensing,Chen2017SLRsensing,Danilov2018SLRsensing}.

Recently, Zhou and Odom demonstrated a new type of subradiant plasmon --- out-of-plane lattice plasmon (OLP) \cite{Odom2011OLPLR}, and found that it is a leaky surface Bloch mode \cite{Odom2012OLPLR}. Compared with ILPs, OLPs are shown to have easier tunability \cite{Odom2011OLPLR}, lower thresholds for surface plasmon amplification \cite{Ding2014OLPlasing}, and much higher quality factors \cite{Boyd2016OLPLR}. This is because OLPs have stronger inter-particle coupling \cite{Odom2011OLPLR,Boyd2016OLPLR}, which suppresses radiative decay, traps light in the plane of the array and strongly localizes fields around each nanoparticle \cite{Odom2011OLPLR}. In order to achieve strong OLPs, Zhou and Odom showed that the height of the nanoparticles should be large ($\geq 100$ nm), and that the incident light should be oblique and TM polarized \cite{Odom2011OLPLR}. OLPs cannot be excited with TE polarized light because there is no out-of-plane electric field component of the incident light \cite{Odom2011OLPLR}. In addition, Huttunen {\sl et al.} found that hexagonal arrays are better than square arrays \cite{Boyd2016OLPLR}. However, the effects of the array period and the dielectric environment on OLPs have not been fully understood yet, hampering the engineering and application of OLPs. For example, although all the above work on OLPs \cite{Odom2011OLPLR,Odom2012OLPLR,Ding2014OLPlasing,Boyd2016OLPLR} adopted homogeneous dielectric environments, in a recent review Kravets {\sl et al.} surmised that for OLPs, the refractive index mismatch between substrate and superstrate is less important since the electric fields associated with the interaction between neighboring particles are primarily in the superstrate \cite{Grigorenko2018SLRreview}.

Here, we report the necessary conditions for metallic nanoparticle arrays to achieve OLPs. We will systematically investigate the effects of key parameters on OLPs, including the array period and the dielectric environment, as well as the nanoparticle height and the angle of incidence. We will examine whether OLPs can be supported in a highly asymmetric dielectric environment. We will also provide intuitive theoretical explanations to our simulation results, unveiling the roles of different parameters. We expect this work will deepen our understanding of OLPs and bring them closer to practical applications in nanolasers, nonlinear optics and sensing.

\section{Simulation setup}
\begin{figure}[!b]
\centering
\includegraphics[width=0.6\linewidth]{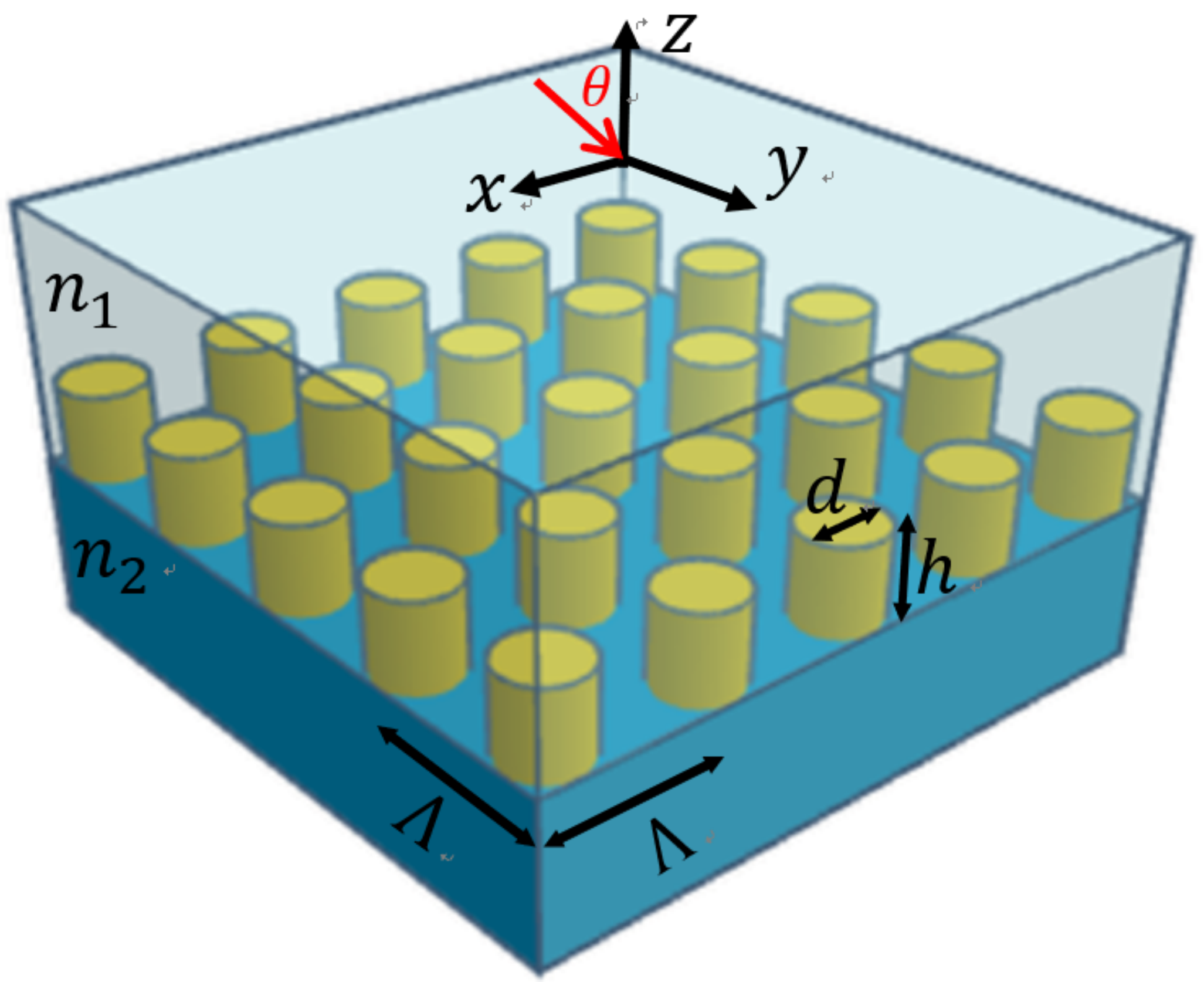}
\caption{Schematic of metallic nanoparticles patterned in a square array. $\Lambda$ is the period in both $x$ and $y$ directions, $h$ and $d$ are the height and diameter of the nanoparticle, respectively, and $n_1$ and $n_2$ are the refractive indices of the thick-enough superstrate and the substrate, respectively. The structure is illuminated by TM-polarized plane waves (magnetic field is parallel to the $y$ direction) with incidence angle $\theta$ in air. The $x-z$ plane is taken to be the plane of incidence.}
\label{fig:Schem}
\end{figure}

Figure \ref{fig:Schem} illustrates the metallic nanoparticles patterned in a square array with period $\Lambda$ in both the $x$ and $y$ directions. Each nanoparticle is cylindrical with height $h$ and diameter $d$. The nanoparticle array is fabricated on a dielectric substrate with refractive index $n_{2}$, and is later covered by a thick-enough superstrate with refractive index $n_{1}$. The structure is illuminated by plane waves at oblique incidence, the plane of which is aligned with the axis of the grating (the $x-z$ plane). Note that here $\theta$ is the angle of incidence in air so as to be consistent with the incidence angle in measurements \cite{Odom2011OLPLR}.

In order to calculate the reflection, transmission and absorption spectra of such a system, we adopt the rigorous coupled wave analysis (RCWA) method, which is very powerful for the analysis and design of periodic structures such as gratings and metamaterials. Calculations were performed using the open-source RCWA software RETICOLO, which was developed by Hugonin and Lalanne \cite{hugonin2005reticolo}. In order to plot the fields at specific wavelengths, we used the commercial finite-difference time-domain (FDTD) software Lumerical FDTD Solutions. 

With the calculated total reflection $R$ and total transmission $T$, we can then calculate the scattering, extinction and absorption cross sections defined as:
\begin{equation}
\label{eq:CrossSec}
\begin{aligned}
\sigma_{\rm sca} &\equiv R \Lambda^{2} \cos\theta, \\
\sigma_{\rm ext} &\equiv (1-T)\Lambda^{2} \cos\theta, \\
\sigma_{\rm abs} &\equiv (1-T-R)\Lambda^{2} \cos\theta.
\end{aligned}
\end{equation}

For all the calculations in this work, we use gold as the material of metallic nanoparticles with refractive index data in the visible and near-infrared regime tabulated in \cite{JC1972NK}. Unless otherwise specified, the incident light is TM polarized with the magnetic field parallel to the $y$ direction, the metallic nanoparticles have a height of $h=100$ nm, a diameter of $d=160$ nm, a period of $\Lambda=400$ nm, the angle of incidence is $\theta=15^{\rm o}$, and the dielectric environment is homogeneous with $n_1=n_2=1.52$.

\section{Results and discussion}
\subsection{Effect of array period}
\begin{figure}[!b]
\centering
\includegraphics[width=\linewidth]{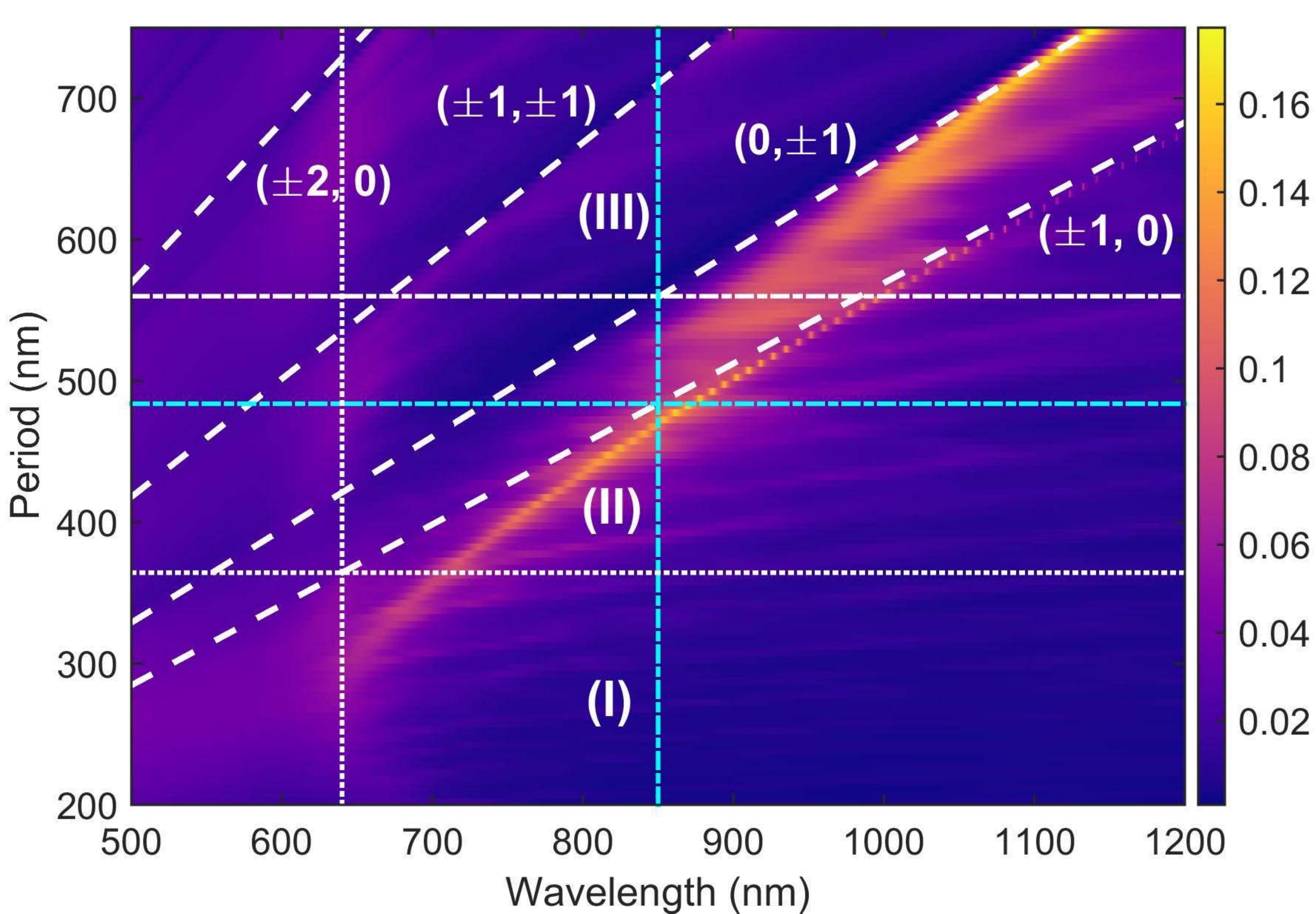}
\caption{Absorption cross section $\sigma_{\rm abs}$ (in unit of $\mu{\rm m}^2$) {\sl versus} period and wavelength. The white dashed lines denote different orders of RAs. The white and blue vertical lines indicate the out-of-plane ( $\lambda_{\rm oLSPR}\approx640$ nm) and in-plane ($\lambda_{\rm iLSPR}=850$ nm) LSPR wavelengths of individual nanoparticles, respectively. From bottom to top, the horizontal lines are the periods for achieving $\lambda_{\pm1,0}=\lambda_{\rm oLSPR}$, $\lambda_{\pm1,0}=\lambda_{\rm iLSPR}$, and $\lambda_{0,\pm1}=\lambda_{\rm iLSPR}$. The white horizontal lines divide the figure into three regions denoted as Regions (I), (II) and (III).}
\label{fig:AbsTM}
\end{figure}
We first investigate the effect of the array period on OLPs, which has not been discussed in the literature yet. Figure \ref{fig:AbsTM} shows the calculated absorption cross section  as a function of the wavelength and the period. To understand the behavior, we also superimpose the positions of the Rayleigh anomaly (RA), which is associated with light diffracted parallel to the in-plane surface. RAs occur when the period and wavelength satisfy \cite{Odom2009RA}
\begin{equation}
\lambda_{p,q} = \dfrac{\Lambda}{p^{2}+q^{2}} \left(\sqrt{(p^2+q^2)n_2^2 - q^2 \sin^2 \theta} \pm p \sin\theta \right),
\label{eq:absCross}
\end{equation}
where integer index pairs $(p,q)$ denote the order of RAs.

\begin{figure}[!tb]
\centering
\includegraphics[width=0.8\linewidth]{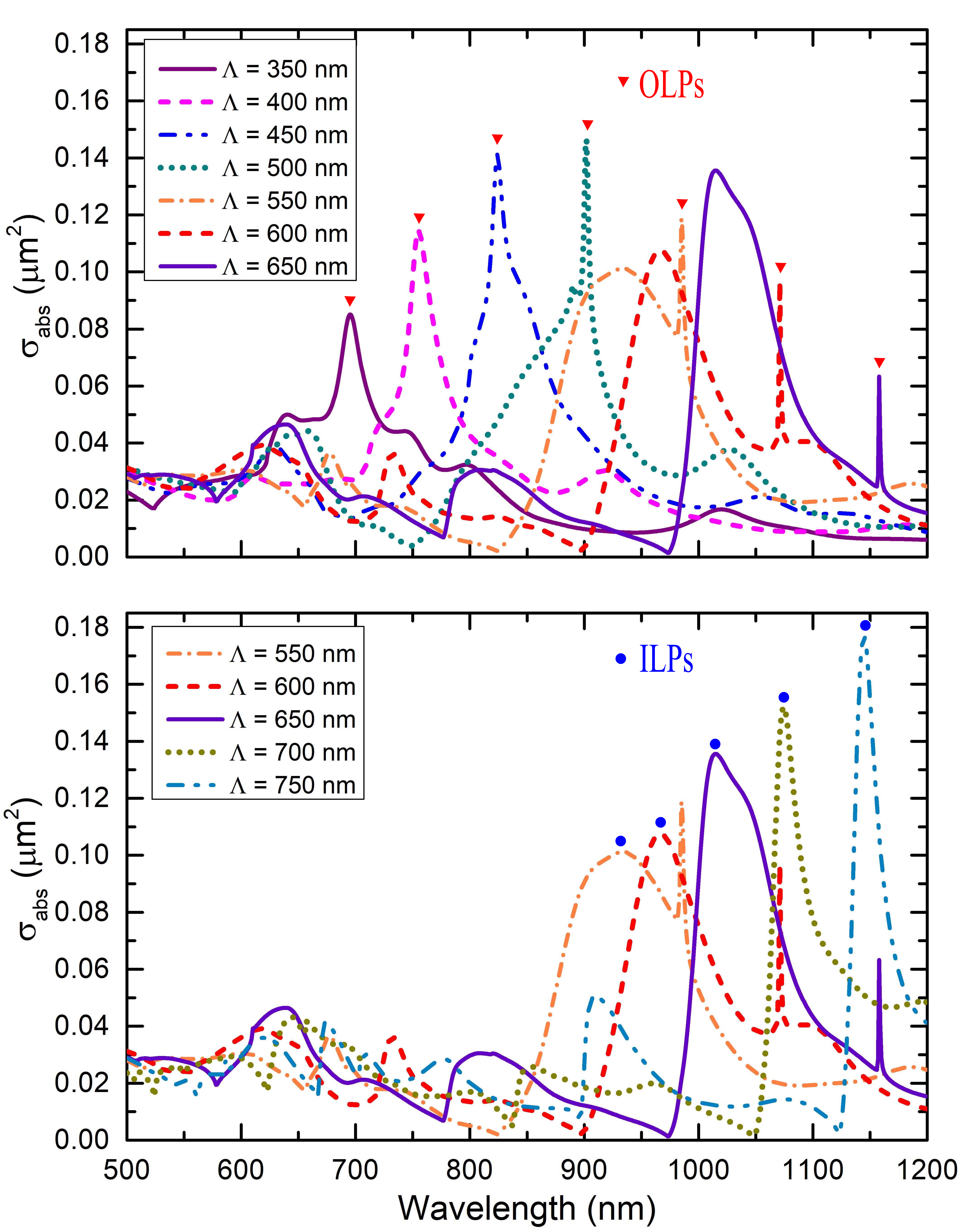}
\caption{Absorption cross section $\sigma_{\rm abs}$ (in units of $\mu{\rm m}^2$) {\sl versus} wavelength for different periods. The resonance peaks of OLPs and ILPs are indicated by red triangles and blue circles, respectively.}
\label{fig:AbsTMWv}
\end{figure}

We find that OLPs with extremely narrow linewidth and large absorption cross section occur only when the period is properly selected from a limited range, $360\,{\rm nm}\lessapprox \Lambda \lessapprox 560\,{\rm nm}$, which is denoted as Region (II) in Fig.~\ref{eq:absCross}. As the period increases within this region, the absorption cross section of OLPs first increases and then decreases, and the linewidth keeps narrowing down. The maximum absorption cross section is achieved when $\lambda_{\pm1,0}\approx\lambda_{\rm iLSPR}$. In this case, out-of-plane dipolar interactions between nanoparticles are well coupled with the in-plane LSPR of individual gold nanoparticles (located around 850 nm \cite{Odom2011OLPLR}), leading to a remarkable narrow absorption peak near $\lambda_{\rm iLSPR}$. Fig.~\ref{fig:AbsTMWv} shows that when the array period is tuned away, the out-of-plane diffraction spectrally deviates from the in-plane LSPR, resulting in reduced absorption cross section. The lower diffraction edge can be roughly determined by the out-of-plane LSPR of individual nanoparticles ($\sim640$ nm) through $\lambda_{\pm1,0}=\lambda_{\rm oLSPR}$, corresponding to $\Lambda\approx 360$ nm, as shown by the white horizontal dotted line in Fig.~\ref{fig:AbsTM}. The upper diffraction edge can be approximately decided by $\lambda_{0,\pm1}=\lambda_{\rm iLSPR}$, corresponding to $\Lambda\approx 560$ nm, as shown by the white horizontal dashed line in Fig.~\ref{fig:AbsTM}.

As the period further increases to $\Lambda \gtrapprox 560$ nm, which is denoted as Region (III) in Fig.~\ref{fig:AbsTM}, ILPs with much wider linewidth appear, and show increasing absorption cross section and decreasing linewidth, as shown by Figs.~\ref{fig:AbsTM} and \ref{fig:AbsTMWv}. OLPs gradually disappear, whereas ILPs gradually dominate and converge to the $(0,\pm 1)$ RA. Note that the resonant wavelength of ILPs depends on the period, and the linewidths are much smaller than that of the in-plane LSPR of individual nanoparticles ($>300$ nm \cite{Odom2011OLPLR}).

For very small periods in Region (I) in Fig.~\ref{fig:AbsTM}, {\sl i.e.}, $\Lambda \lessapprox 360\,{\rm nm}$, the absorption cross section is very weak although broadband. For such small periods that are close to the diameter of the cylindrical gold nanoparticle, the plasmonic structure behaves like a bulk metal with small dielectric inclusions \cite{Gunnarsson2005confined}. In this scenario, the near-field coupling between adjacent nanoparticles is very strong, resulting in enhanced localized gap plasmons.

\begin{figure}[!tb]
\centering
\includegraphics[width=\linewidth]{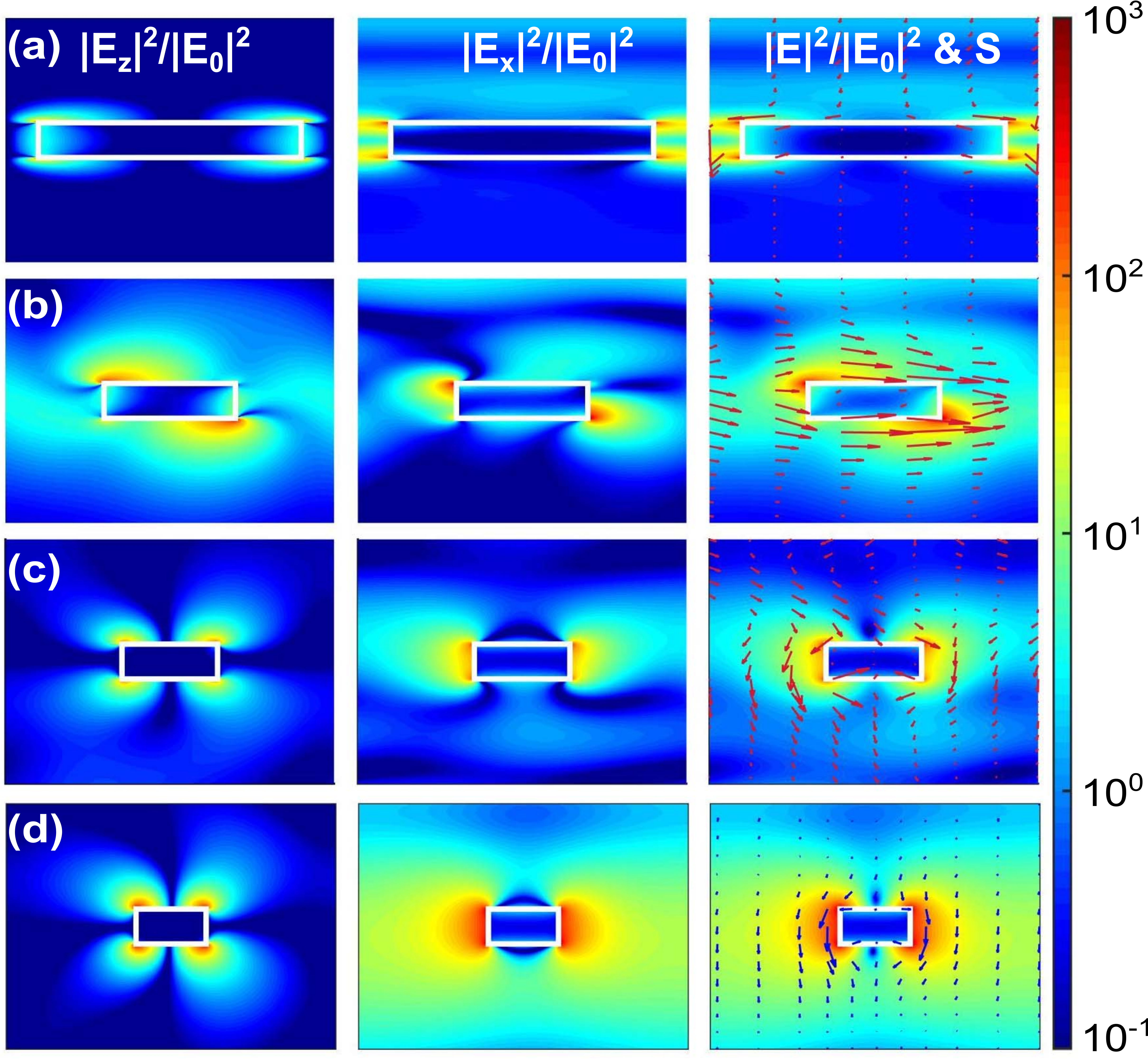}
\caption{Electric field intensity distributions (in color) $|E_z|^2$ (left column), $|E_x|^2$ (center column), $|E|^2$ (right column), and energy flow (Poynting vector map, in arrows, right column) of (a) the array LSPR for $\Lambda=200$ nm, $\lambda=640$ nm, (b) of the narrow OLP at $\Lambda=400$ nm, $\lambda=758$ nm, and of the broad ILPs for (c) $\Lambda=550$ nm, $\lambda=935$ nm, and for (d) $\Lambda=750$ nm, $\lambda=1145$ nm. The white rectangles outline the nanoparticle with height 100 nm and diameter 160 nm. The cross sections are aligned with the axis of the grating, {\sl i.e.}, the $x-z$ plane, the total $x$ width is the period $\Lambda$, and the total $z$ height is 700 nm. The calculations were performed with unitary electric field intensity for the incident light $|E_0|^2$.}
\label{fig:FieldsVsP}
\end{figure}

Figure~\ref{fig:FieldsVsP}(a) shows that the electric field of the localized plasmon mode for $\Lambda=200$ nm and $\lambda=640$ nm is dominated by the $x$ component, and is mainly confined to the gap due to strong near-field coupling between neighboring nanoparticles. Fig.~\ref{fig:FieldsVsP}(b) shows that the electric field of the OLP for $\Lambda=400$ nm and $\lambda=758$ nm is dominated by its out-of-plane component $E_z$. The Poynting vector map shows that the energy flux propagates along $x$ and is concentrated above and below the plane of the two-dimensional nanoparticle array. Therefore, the out-of-plane lattice plasmon mode is an electromagnetic wave that propagates in-plane but polarizes out-of-plane. Fig.~\ref{fig:FieldsVsP}(c) shows that the electric field of the ILP is dominated by its in-plane component $E_x$, and the energy flux circulates around the nanoparticles. As the period increases from 550 nm to 750 nm, as shown by Fig.~\ref{fig:FieldsVsP}(c)(d), the coupling between in-plane  nanoparticle dipolar moments is greatly enhanced, resulting in increased absorption cross section and narrowed linewidth of the ILP (see Fig.~\ref{fig:AbsTMWv}).

\begin{table}
\centering
\caption{Period dependent dominant effects for TM polarized light with oblique angle of incidence. $\lambda_{\rm iLSPR}$ and $\lambda_{\rm oLSPR}$ means the resonant wavelengths of the in-plane and out-of-plane LSPRs of individual nanoparticles, respectively.}
\begin{tabular}{cc}
\hline
Period & Effect \\
\hline
$\Lambda \lessapprox  \dfrac{\lambda_{\rm oLSPR}}{{(n_2^2-\sin^2\theta)^{1/2}}+\sin\theta} $ & Near-field coupling \\
$ \dfrac{\lambda_{\rm oLSPR}}{{(n_2^2-\sin^2\theta)^{1/2}}+\sin\theta} \lessapprox \Lambda \lessapprox \dfrac{\lambda_{\rm iLSPR}}{n_2}$ & OLPs \\
$\Lambda \gtrapprox \dfrac{\lambda_{\rm iLSPR}}{n_2} $ & ILPs\\
\hline
\end{tabular}
  \label{tab:summary}
\end{table}

Therefore, given the individual metallic nanoparticles, the array period determines both the strength and the spectral position of the absorption cross section of the OLPs. Table~\ref{tab:summary} summarizes the dominant effects for different periods in Regions (I)-(III) in Fig.~\ref{fig:AbsTM}. For $\Lambda \lessapprox \lambda_{\rm oLSPR}/[{(n_2^2-\sin^2\theta)^{1/2}}+\sin\theta]$, the inter-particle gap is small enough that the array LSPR dominates; for $\Lambda \gtrapprox \lambda_{\rm iLSPR}/n_2 $, ILPs become dominant; only for periods that are properly designed within $\lambda_{\rm oLSPR}/[{(n_2^2-\sin^2\theta)^{1/2}}+\sin\theta] \lessapprox \Lambda \lessapprox \lambda_{\rm iLSPR}/n_2$, OLPs with extremely narrow linewidths and large absorption cross sections can be observed. The strongest OLP is achieved when the array period is around $\Lambda = \lambda_{\rm iLSPR}/[{(n_2^2-\sin^2\theta)^{1/2}}+\sin\theta]$, which satisfies $\lambda_{\pm1,0} = \lambda_{\rm iLSPR}$.

While we mostly concentrated on the TM polarization, which has the out-of-plane electric component ($E_z$) of the incident light, we also found that TE polarized incident light, which has only the in-plane electric component ($E_y$), cannot excite OLPs, consistent with the results obtained by Zhou and Odom \cite{Odom2011OLPLR}.

\subsection{Effect of dielectric environment}
Previously, we have taken the dielectric environment to be homogeneous, which simplifies theoretical analysis and simulations. In experiments, however, inhomogeneous dielectric environments are easier to implement since the refractive indices of the superstrate and of the substrate are usually different. It is known that the optimal ILPs is achieved when the superstrate and substrate are index matched \cite{Abajo2010ILPsub,Khlopin2017IPSLRAuAl,Odom2018SLRreview}. Here we show that a homogeneous dielectric environment is also vital for achieving OLPs. 

\begin{figure}[!tb]
\centering
\includegraphics[width=\linewidth]{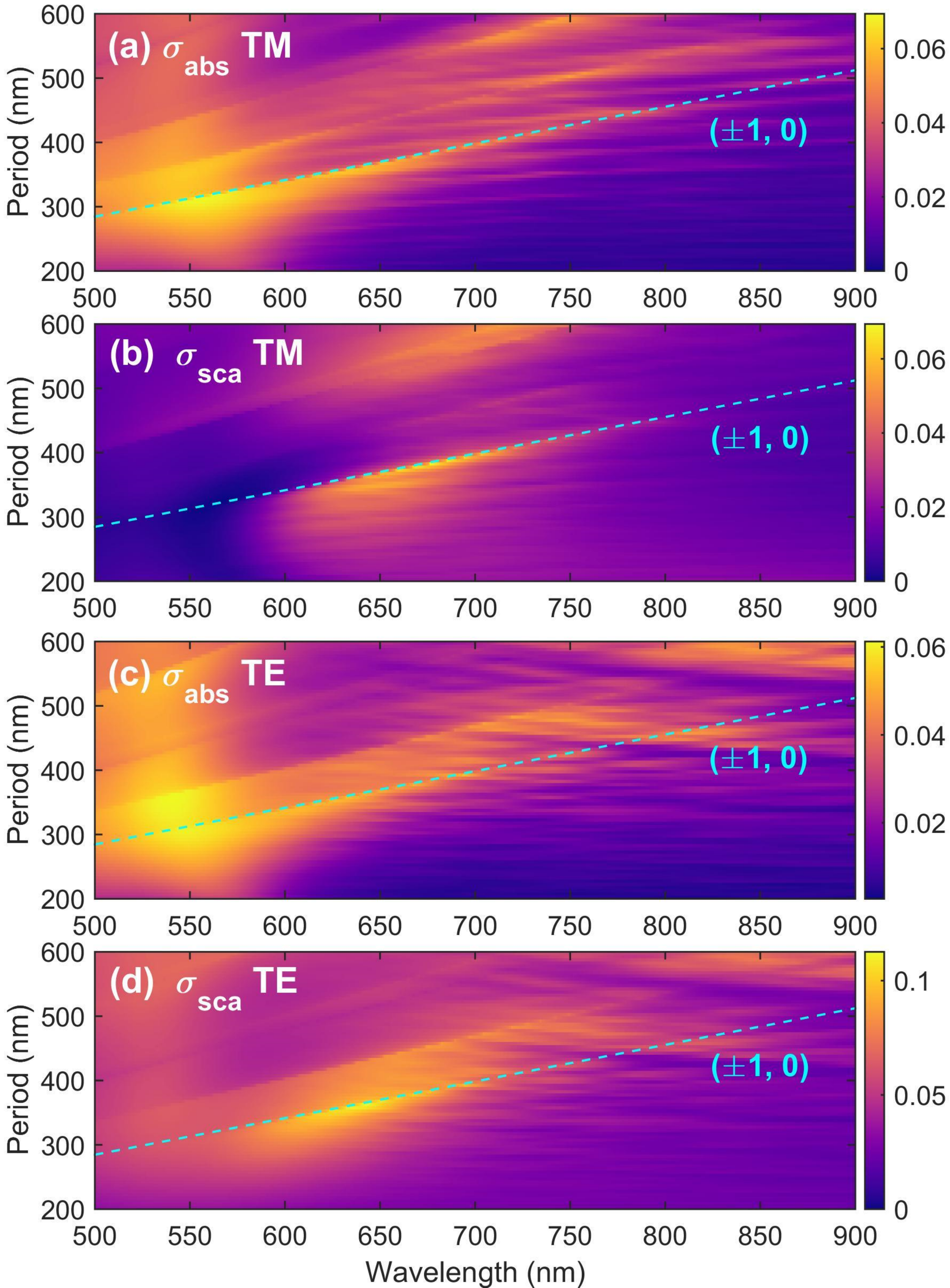}
\caption{(a)(c) Absroption and (b)(d) scattering cross sections (in unit of $\mu {\rm m}^2$) {\sl versus} period and wavelength for inhomogeneous dielectric environment: $n_1=1.0$ (air) and $n_2=1.52$. (a) and (b) are for TM polarized incidence, and (c) and (d) are for TE polarized incidence. The blue dsshed lines denote $(\pm1,0)$ order RAs.}
\label{fig:AbsTMAir}
\end{figure}

\begin{figure}[!tb]
\centering
\includegraphics[width=\linewidth]{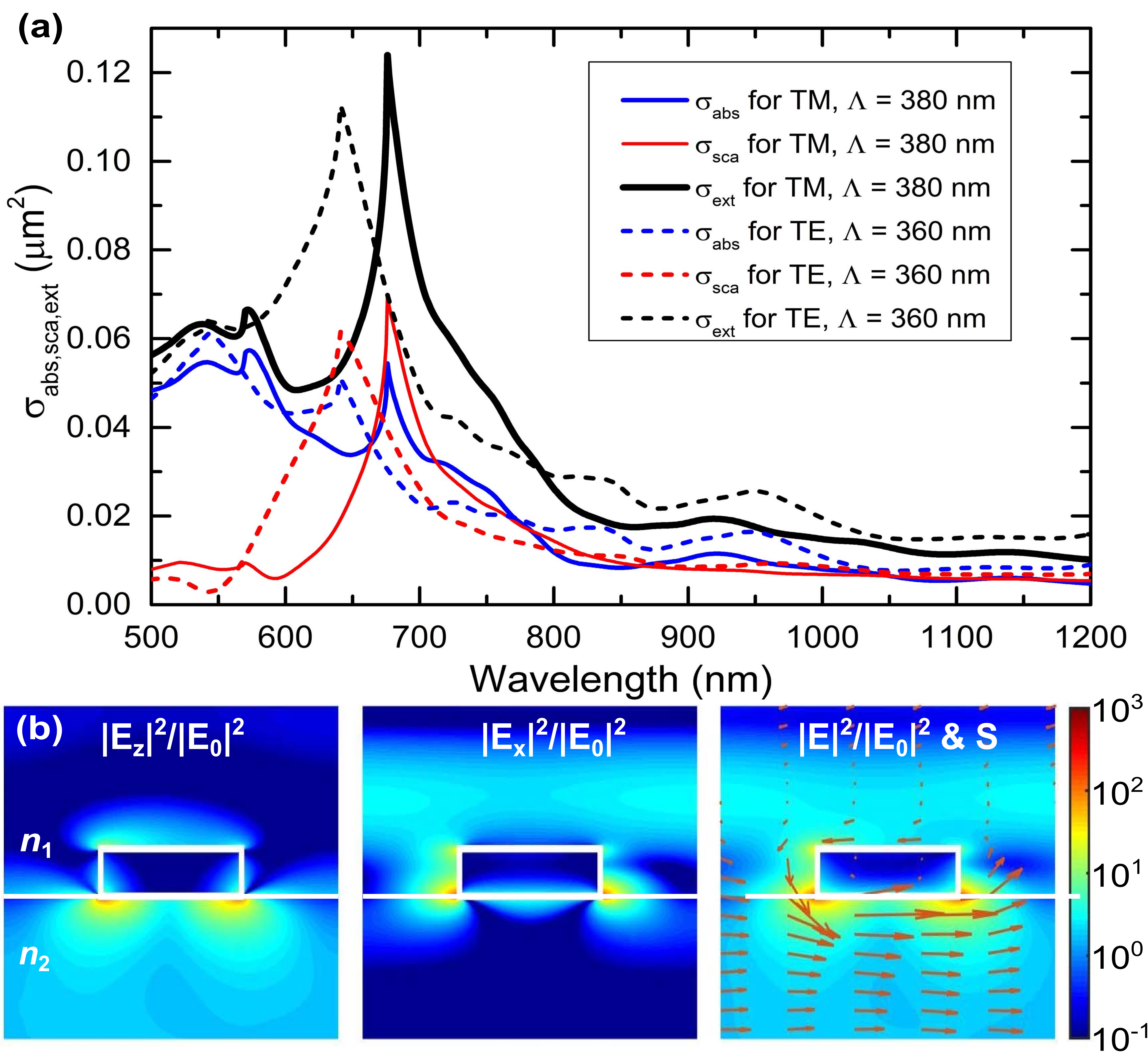}
\caption{(a) Absorption (blue), scattering (red) and extinction (black) cross section spectra for $\Lambda = 380~{\rm nm}$ under TM polarized oblique incidence (solid lines) and for $\Lambda = 360~{\rm nm}$ under TE polarized oblique incidence (dashed lines). (b) Electric field intensity distributions (in color) and energy flow (in arrows) for $\Lambda = 380~{\rm nm}$ under TM polarized oblique incidence at $\lambda=676~{\rm nm}$. The calculations were performed for nanoparticle array in inhomogeneous dielectric environment: $n_1=1.0$ (air) and $n_2=1.52$ (polymer).}
\label{fig:FieldAir}
\end{figure}

We consider a large refractive index difference between the superstrate and the substrate: $n_{1}=1.0$ (air) and $n_2=1.52$. For such an inhomogeneous dielectric environment and TM polarized oblique incidence, Fig.~\ref{fig:AbsTMAir}(a)(b) shows that when the period is properly selected (around 380 nm, also around the $(\pm1,0)$ order RA line), both the absorption and the scattering cross section spectra show a narrow peak. The resulting extinction cross section spectra ($\sigma_{\rm ext} = \sigma_{\rm abs} + \sigma_{\rm sca}$), as shown in Fig.~\ref{fig:FieldAir}(a) shows a peak at $\lambda = 676$ nm with a linewdith of $\sim23$ nm, corresponding to a quality factor of 29. This quality factor is larger than that obtained for a 200 nm tall gold nanoparticle array in a asymmetric air/glass environment ($Q\approx 19$), as reported by Thackray {\sl et al.} \cite{Grigorenko2014OLPLR}.

In order to understand such a narrow collective plasmon resonance, which was attributed to the out-of-plane coupling of the localized surface
plasmon resonances of the individual nanostructures \cite{Grigorenko2014OLPLR}, we calculated the corresponding electric fields and the Poynting vector map, as shown in Fig.~\ref{fig:FieldAir}(b). The results show that the in-plane dipole interactions dominate, the circulating energy flows around the nanoparticle, and the energy flux propagates along the $x$ direction and is concentrated below the nanoparticle array. This suggests that this collective plasmon resonance is an ILP rather than an OLP. To further confirm that an ILP is excited in this scenario, we also calculated the results for the TE polarization, as shown in Fig.~\ref{fig:AbsTMAir}(c)(d). Comparing Figs.~\ref{fig:AbsTMAir} and \ref{fig:FieldAir}(a), it is evident that narrow collective plasmon resonances with similar linewidths can be excited for both TE and TM polarization. Additionally, we find that the interactions betsween neighboring particles are primarily in the substrate, rather than in the superstrate. These results are opposite to the hypothesis by Kravets {\sl et al.}, who surmised that for OLPs, which are supported in tall nanoparticles and are excited under oblique incidence, the refractive index mismatch between substrate and superstrate is less important since the electric fields associated with the interactions between neighboring particles are primarily in the superstrate \cite{Grigorenko2018SLRreview}.

Therefore, our results suggest that OLPs require homogeneous dielectric environment even more than ILPs. It has been shown that ILPs supported by short (usually 50 nm tall) metal nanoparticle arrays prefer homogeneous dielectric environments \cite{Khlopin2017IPSLRAuAl}, and ILPs supported by tall (for example 200 nm tall) metal nanoparticle arrays can have narrow linewidths for an environment as asymmetric as the air/glass environment \cite{Grigorenko2014OLPLR}. This is because for inhomogeneous dielectric environments, the electric field distribution is highly asymmetric and most of the energy is confined to the high-index substrate; in-plane dipolar interactions can interact strongly through the RA in the substrate, as shown by Figs.~\ref{fig:AbsTMAir} and \ref{fig:FieldAir}(b); however, out-of-plane dipolar interactions are mainly in the superstrate (air in this case), leading to very weak out-of-plane inter-particle interactions due to the weak field in the superstrate.

\subsection{Effects of height and incidence angle}
\begin{figure}[!tb]
\centering
\includegraphics[width=0.8\linewidth]{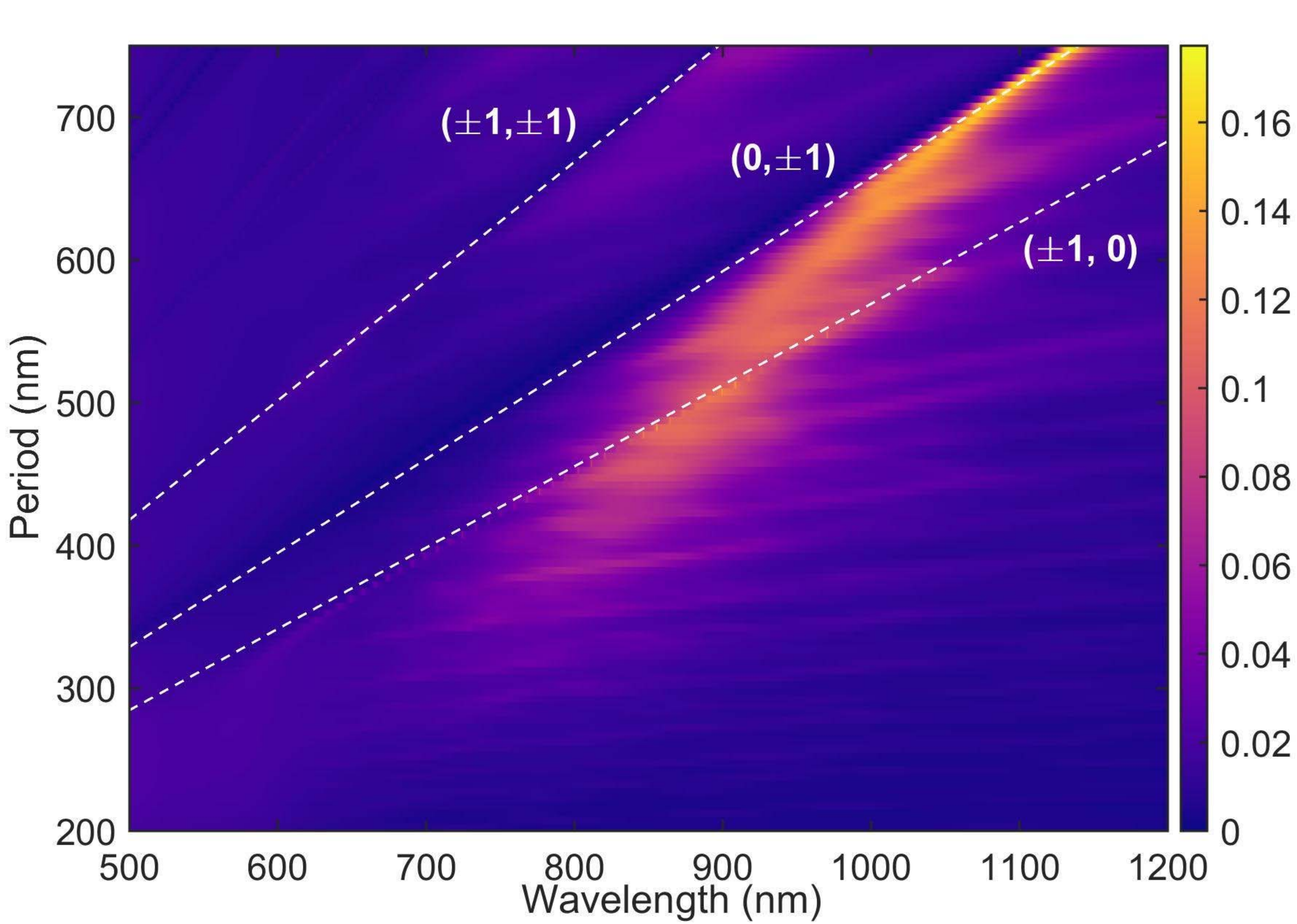}
\caption{Similar to Fig.~\ref{fig:AbsTM}, but calculated with short nanoparticles with height $h=50$ nm.}
\label{fig:AbsvsPWvH50}
\end{figure}

Zhou and Odom found that OLPs can be pronounced only for $h \geq 100$ nm given $\theta=15\degree$, and that the absorption cross section is optimized for $\theta=15\degree$ given $h=100$ nm \cite{Odom2011OLPLR}. This finding helps to understand the fact that most work has been focused on ILPs, and has failed to observe OLPs: nanoparticles that are easy to fabricate are too short (usually $h\leq 50$ nm), or incidence that is easy to implement in experiments is normal incidence ({\sl i.e.}, $\theta=0\degree$). Indeed, for short nanoparticles with $h=50$ nm, we explored all the other parameters and found that OLPs are too weak to be observed, but ILPs still exist and eventually converge to the $(0,\pm1)$ order RA, as shown by Fig.~\ref{fig:AbsvsPWvH50}. This is because the height only affects the out-of-plane oscillation \cite{Odom2011OLPLR}, whereas the in-plane oscillation mainly depends on the nanoparticle diameter and the period \cite{Khlopin2017IPSLRAuAl}.

\begin{figure}[!tb]
\centering
\includegraphics[width=0.8\linewidth]{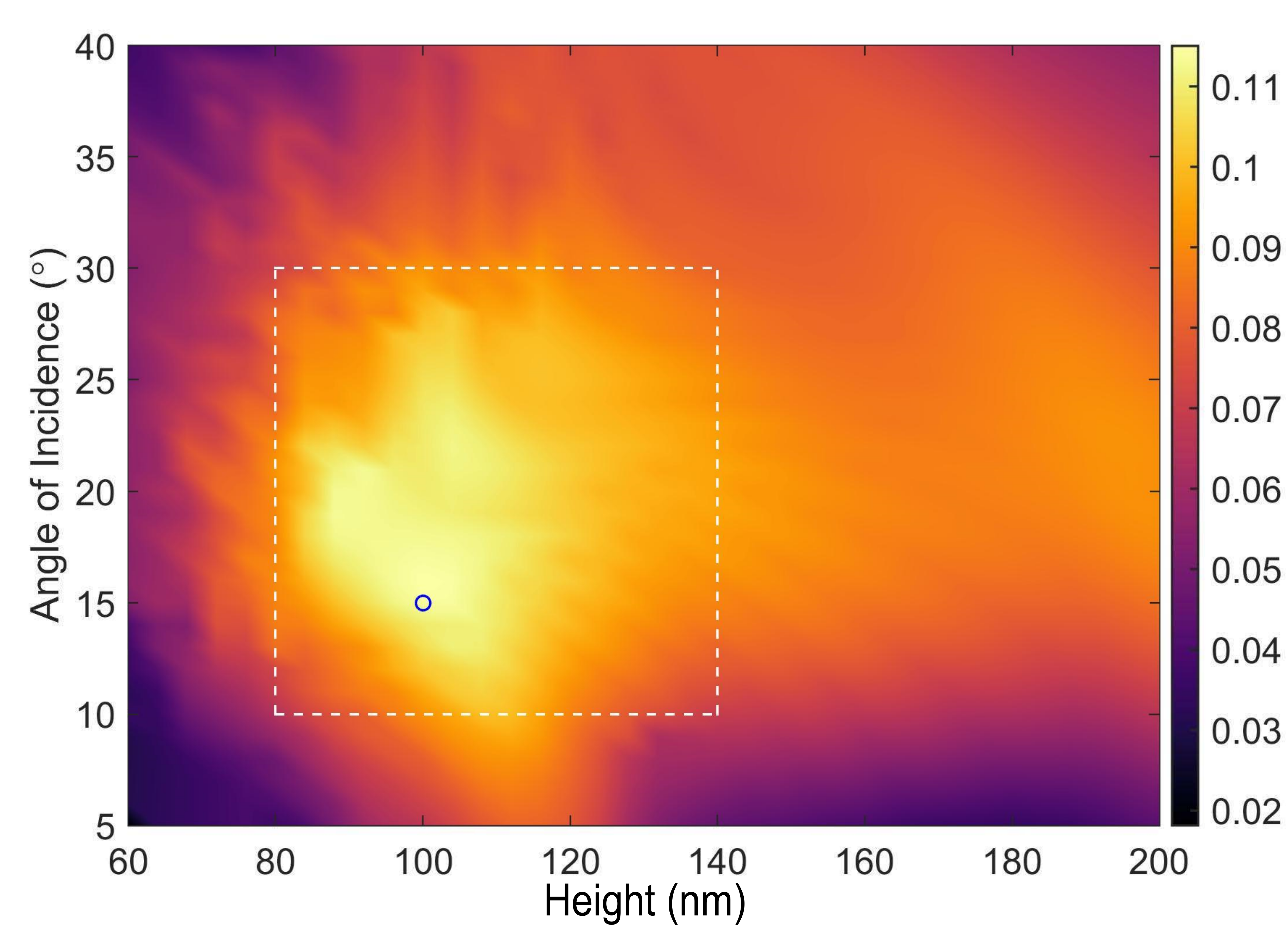}
\caption{Peak absorption cross section $\sigma_{\rm abs,peak}$ (in units of $\mu{\rm m}^2$) {\sl versus} angle of incidence and nanoparticle height. The dashed box indicates the range of nanoparticle heights and angle of incidences for achieving strong OLPs, and the circle indicates the optimal parameters.}
\label{fig:AbsvsHAng}
\end{figure}

Here we systematically study the effects of both the nanoparticle height $h$ and the angle of incidence $\theta$, both of which vary simultaneously (on the contrary, the discussion was performed by varying only one parameter given the other one in \cite{Odom2011OLPLR}). Figure~\ref{fig:AbsvsHAng} shows the peak absorption cross section $\sigma_{\rm abs,peak}$ as functions of both $h$ and $\theta$. Here $\sigma_{\rm abs,peak}$ denotes the peak value of the $\sigma_{\rm abs}$ spectrum, {\sl i.e.}, $\sigma_{\rm abs,peak} = {\rm max} \{\sigma_{\rm abs}(\lambda)\}$. Results show that a large absorption cross section can be achieved only for a small range of nanoparticle heights and angles of incidence: $80~{\rm nm}\lessapprox h\lessapprox 140~{\rm nm}$, and $10^{\circ}\lessapprox\theta\lessapprox 30^{\circ}$, as indicated by the dashed rectangle in the figure. The optimal nanoparticle height and angle of incidence are around $h=100~{\rm nm}$ and $\theta=15^{\circ}$, consistent with Zhou and Odom \cite{Odom2011OLPLR}.

The optimal range for nanoparticle height and angle of incidence can be understood as follows. For too small a $\theta$, the out-of-plane electric field component $|E_0| \sin \theta$ is too weak to excite out-of-plane dipoles. As $\theta$ increases, the out-of-plane electric field component increases, leading to larger $\sigma_{\rm abs}$. However, if $\theta$ is too large, the out-of-plane lattice plasmon spectrally shifts beyond the broad in-plane lattice plasmon \cite{Odom2011OLPLR}. On the other hand, nanoparticles that are too short cannot support out-of-plane dipoles, resulting in small $\sigma_{\rm abs}$ originating from ILPs. As the nanoparticle height increases, out-of-plane dipoles and OLPs can be supported, leading to larger $\sigma_{\rm abs}$. However, if $h$ is too large, the out-of-plane dipole suffers from high Ohmic loss due to the long nanoparticle, which limits the inter-particle coupling and reduces $\sigma_{\rm abs}$ of OLPs. Therefore, there is a compromise for both the angle of incidence and the nanoparticle height. Similar phenomena can also be observed and explained similarly for other periods in Regime (II) in Fig.~\ref{fig:AbsTM}.

\section{Conclusion}
In conclusion, we have investigated the effects of key parameters of metallic nanoparticle arrays, including the array period, the dielectric environment, the nanoparticle height and the angle of incidence, on OLPs. By doing this, we have managed to clarify the necessary conditions for achieving strong OLPs. Our results have shown that in addition to the oblique incidence, the TM polarization, and the large nanoparticle height, which have been pointed out by the literature; other necessary conditions include a properly designed array period within a limited range of $\lambda_{\rm oLSPR}/[{(n_2^2-\sin^2\theta)^{1/2}}+\sin\theta] \lessapprox \Lambda \lessapprox \lambda_{\rm iLSPR}/n_2$, a homogeneous dielectric environment, and optimized pair of the nanoparticle height and the angle of incidence. We also clarify that for tall metal nanoparticles in an environment as asymmetric as air/polymer (with a large refractive index difference 1.0/1.52) environment, it is the ILP rather than the OLP that can be excited. We expect these findings will advance the understanding and the design of OLPs, as well of ILPs in tall metal nanopartical array, and promote OLPs with extremely narrow linewidth and high quality factor for applications in nanolasers, nonlinear optics and ultrasensitive sensing.

\section*{Funding}
Key Research Items from the Ministry of Science and Technology of China (Grant No. 2016YFC1400701); State Key Laboratory of Advanced Optical Communication Systems Networks, China (Grant No. 2019GZKF2).

\bibliographystyle{unsrt}
\bibliography{sample}

\end{document}